\documentclass[aps,prb,reprint,groupedaddress, showkeys, showpacs, superscriptaddress]{revtex4-1}
 
\pdfoutput=1

\usepackage{amsmath}
\usepackage{graphicx}
\usepackage{dcolumn}
\usepackage{verbatim}

\begin{document}

\title{The role of the ligand layer for photoluminescence spectral diffusion of CdSe/ZnS nanoparticles}

\author{Daniel Braam}
\email{Daniel.Braam@uni-due.de}
\affiliation{Fakult\"at f\"ur Physik and CENIDE, Universit\"at Duisburg-Essen, Lotharstra\ss{}e 1, 47057 Duisburg, Germany}
\author{Andreas M\"olleken}
\affiliation{Fakult\"at f\"ur Physik and CENIDE, Universit\"at Duisburg-Essen, Lotharstra\ss{}e 1, 47057 Duisburg, Germany}
\author{G\"unther M. Prinz}
\affiliation{Fakult\"at f\"ur Physik and CENIDE, Universit\"at Duisburg-Essen, Lotharstra\ss{}e 1, 47057 Duisburg, Germany}
\author{Christian Notthoff}
\affiliation{Nanoparticle Process Technology and CENIDE, Universit\"at Duisburg-Essen, Lotharstra\ss{}e 1, 47057 Duisburg, Germany}
\author{Martin Geller}
\affiliation{Fakult\"at f\"ur Physik and CENIDE, Universit\"at Duisburg-Essen, Lotharstra\ss{}e 1, 47057 Duisburg, Germany}
\author{Axel Lorke}
\affiliation{Fakult\"at f\"ur Physik and CENIDE, Universit\"at Duisburg-Essen, Lotharstra\ss{}e 1, 47057 Duisburg, Germany}

\begin{abstract}

The time-resolved photoluminescence (PL) characteristics of single CdSe/ZnS nanoparticles, embedded in a PMMA layer is studied at room temperature. We observe a strong spectral jitter of up to 55\,meV, which is correlated with a change in the observed linewidth. We evaluate this correlation effect using a simple model, based on the quantum confined Stark effect induced by a diffusing charge in the vicinity of the nanoparticle. This allows us to derive a mean distance between the center of the particle and the diffusing charge of approximately 3.3\,nm on average, as well as a mean charge carrier displacement within the integration time. The distances are larger than the combined radius of particle core and shell of about 3\,nm, but smaller than the overall radius of 5\,nm including ligands. These results are reproducible, even for particles which exhibit strong blueing, with shifts of up to 150\,meV. Both the statistics and its independence of core-shell alterations lead us to conclude that the charge causing the spectral jitter is situated in the ligands.  

\end{abstract}



\maketitle

%Introduction
\section{Introduction}
Semiconductor nanoparticles with their zero-dimensional density of states are a highly interesting material system both for research and industry. In contrast to epitaxially grown self-assembled quantum dots  \cite{Bimberg1999}, nanoparticles can be fabricated in large quantities from solution \cite{MURRAY1993} or from the gas phase \cite{LorkeBook2012} using different material combinations. Their high quantum efficiency and their size-tunable emission energy, ranging from the infrared to the visible spectrum, make them promising starting materials for applications in many fields, such as quantum dot lasers \cite{Klimov2000}, biological markers \cite{Bruchez1998}, displays \cite{Kim2011}, and multi-exciton-generation solar cells \cite{Chang2013}.

However, two phenomena may adversely affect their optical properties: (i) The fluorescence intermittency (referred to as "blinking"), observed as a random switching between an emitting "on-state" and a non-emitting "off-state" \cite{Nirmal1996,Frantsuzov2008,Efros2008}. (ii) The spectral diffusion (referred to as "jitter"), observed as a random spectral shift of the emission line \cite{Neuhauser2000,Fernee2010}. The fluorescence intermittency is most likely caused by an excess charge in the nanoparticle, which enables an Auger process and thus leads to a fast non-radiative decay \cite{Efros2008}. Blinking can be suppressed by changing the confinement potential \cite{Mahler2008,Chen2008,Wang2009} or coupling the exciton transition to a metal interface \cite{Kulakovich2002,Jares-Erijman2003,Govorov2006,Pompa2006,Chan2009}. The  spectral jitter is commonly attributed to charge diffusion in the vicinity of the nanoparticle, causing a randomly varying quantum confined Stark shift \cite{Empedocles1999}. However, little is known about the location of the diffusing charge, which is an essential question in order to reduce this source of spectral impurity. On single elongated CdSe nanocrystals, M\"uller et al. \cite{Muller2004,Muller2005} have observed a charge meandering on the surface of the particle. From a study on how an organic matrix, surrounding CdSe/CdS/ZnS nanocrystals, will affect their spectral diffusion, Gomez at al. \cite{Gomez2006} concluded that the diffusing charge is located either at the surface of the NCs or directly at the core-shell interface. 

To further elucidate this question, we use the well-known correlation between energy shift and linewidth  \cite{Empedocles1997,Muller2004,Muller2005} in combination with a simple Coulomb model to obtain statistics of the distance between the  nanoparticle and the diffusing charge. Surprizingly, we find a strongly peaked distribution at a distance, which is larger than the crystalline radius (core plus shell) of the particle, but smaller than the overall radius of about 5\,nm. This strongly suggests that the charge responsible for the spectral jitter is located in the ligand layer, which surrounds each nanoparticle to prevent agglomeration. These results are supported by measurements on particles that are affected by photooxidation in which the core is oxidized and therefore a "blueing" is observed. 

%experimental part
\section{Experimental Details}
The investigated particles consist of a 2\,nm CdSe core and a roughly 1\,nm thick ZnS shell with about 2\,nm of ligands \cite{Evidots}.
They are dispersed in toluene ($C_7H_8$), to which 1\% PMMA (polymethylmetacrylate) has been added as a protective polymer. This solution is spin-coated on metal-coated silicon substrates. To enable single particle spectroscopy, an extremely dilute nanoparticle dispersion is used (about 1\,pmol/ml), which results in much less than 1 particle per $\mu$m$^2$. The nanoparticles are excited nonresonantly with a power density of about 5\,W/cm$^2$ using a 532\,nm frequency doubled Nd:YVO$_4$ laser. The PL emission is collected using a 50x objective (NA=0.5) and detected with a liquid-nitrogen-cooled CCD camera, attached to a 500\,mm spectrometer. The integration time is usually 1--2\,s for a full spectrum with 0.26\,nm (or correspondingly 0.9\,meV) resolution.

\begin{figure*}[htb]
\includegraphics[width=0.8\textwidth]{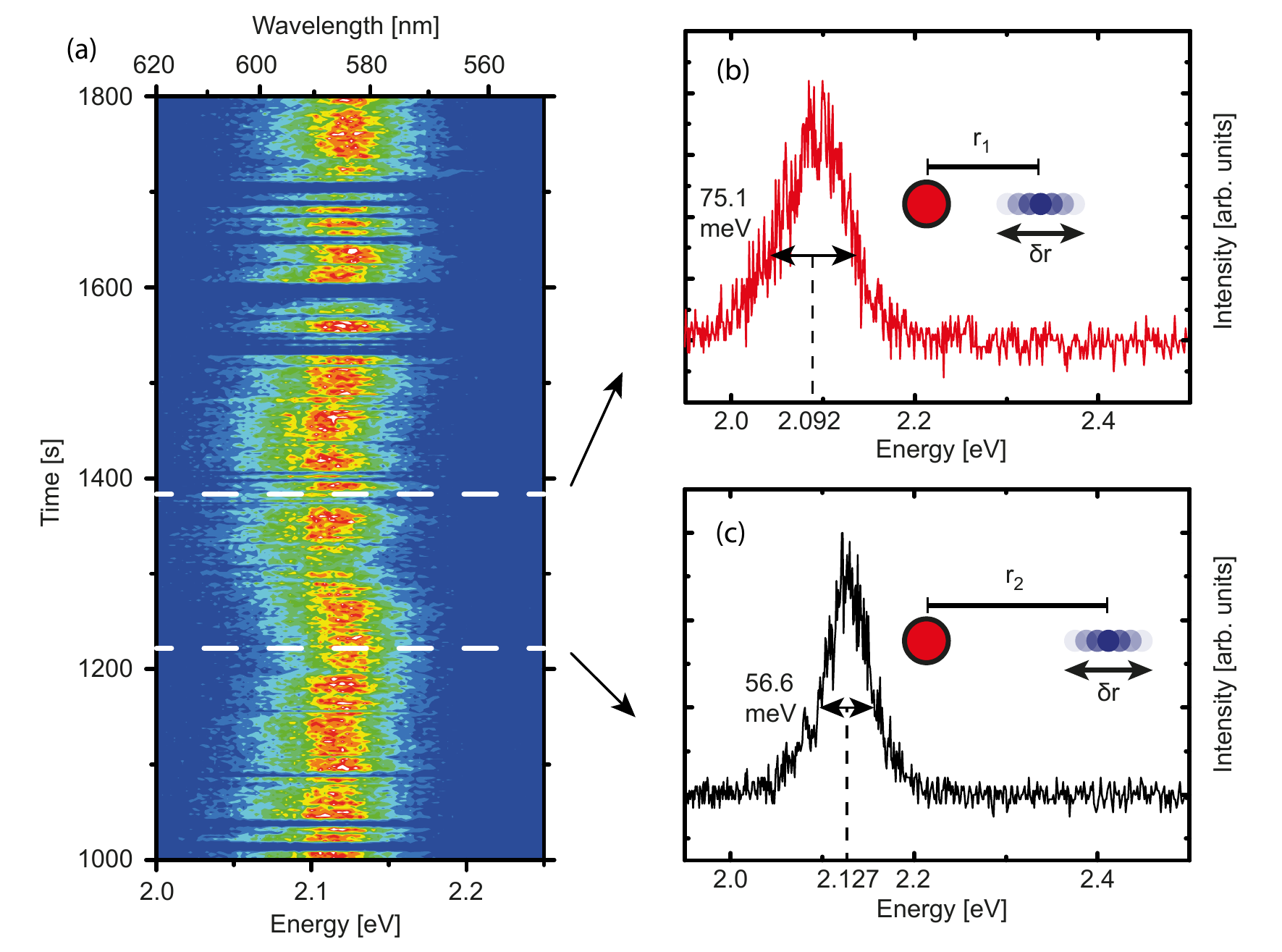}%
\caption{Spectral time evolution of a single CdSe/ZnS nanoparticle (a) with two representative spectra, (b) and (c). Insets illustrate how an electron with distance $r$ from the center of the nanoparticle and constant fluctuation $\delta r$ induces a shift of the emission line to lower energy (a "redshift") and a spectral broadening. The intensity in (a) is colour-coded from blue (low) to red (high).
\label{Spectral_Jitter}
}
\end{figure*}

\begin{figure}[htb]   \includegraphics[width=0.45\textwidth]{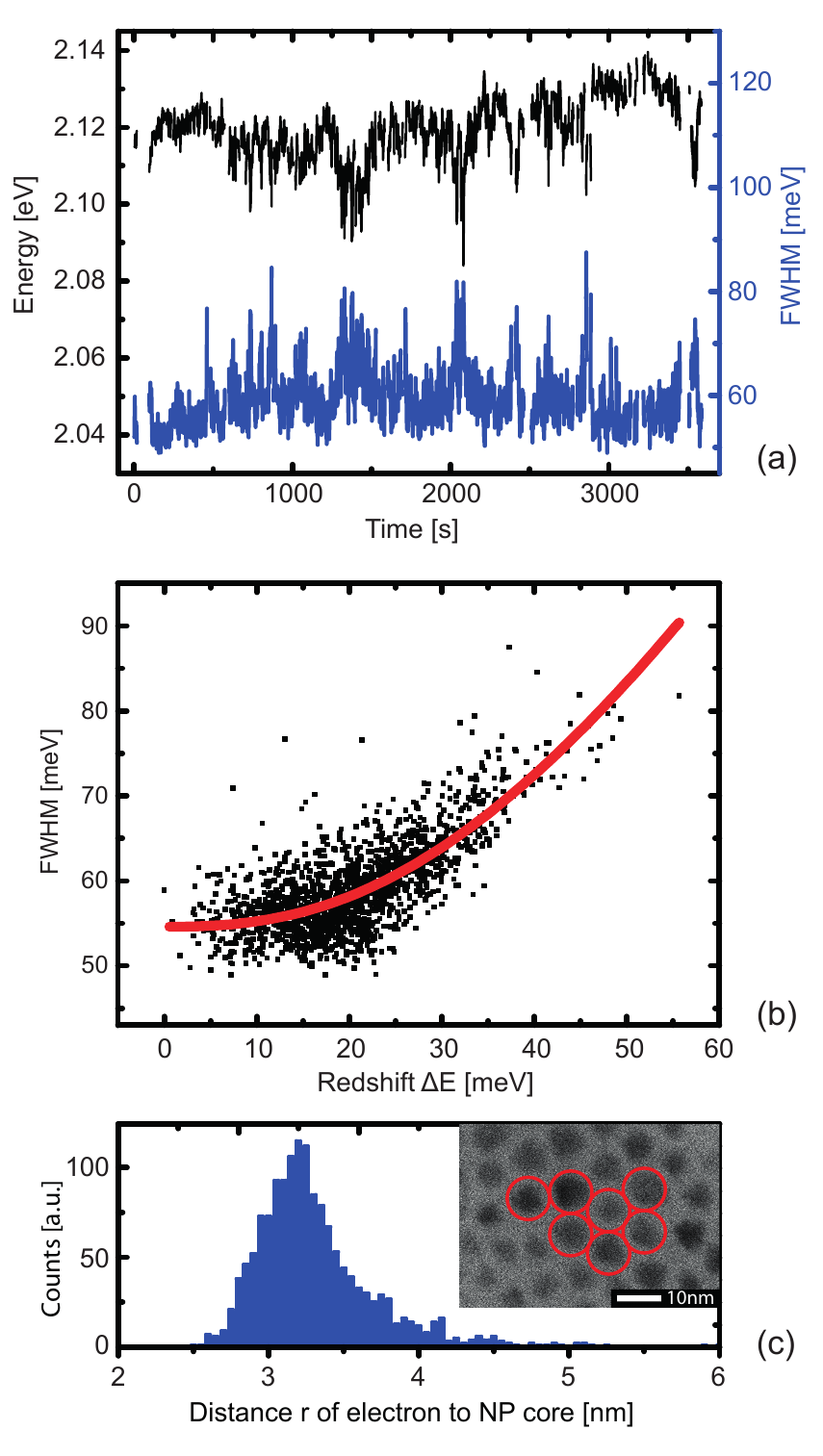}%
   \caption{(a) Time-evolution of the emission peak energy (top black line) and corresponding linewidth (bottom blue line) of a single CdSe/ZnS nanoparticle at room temperature. (b) The FWHM in dependence of the emission redshift shows a superlinear behaviour, which can be fitted using Eqn.~(\ref{eqn_DeltaF}); displayed as a red line. (c) Distribution of the distances $r$, derived from the measured redshifts in (b), calculated by using Eqn.~(\ref{eqn_r}). The inset shows a STEM image of a hexagonal, close-packed lattice of the measured CdSe/ZnS nanoparticles. The circles indicate a 5\,nm radius.
   \label{Energy_and_FWHM_110908_1_Big}
   }
%\end{center}
\end{figure}

\section{Results and Discussion}
Single nanoparticles were identified by their characteristic blinking behavior, and their PL spectra were taken continuously over several hours. A subset of such a data set is shown in Fig.~\ref{Spectral_Jitter}(a). The emission lines were fitted to a Lorentzian function to determine both the spectral peak positions and the full width at half maximum (FWHM, Fig.~\ref{Spectral_Jitter}(b,c)). Figure~\ref{Energy_and_FWHM_110908_1_Big}(a) shows one of more than 70 time traces of PL energies and the corresponding linewidths obtained this way. As observed by Empedocles et al. \cite{Empedocles1997}, the peak position exhibits spectral diffusion with a characteristic asymmetry: Strong shifts are observed mainly towards lower energies. Similarly, also the FWHM fluctuates, with strong shifts mainly towards higher values. Moreover, a correlation between both time traces is apparent, so that the linewidth increases with decreasing peak energy and vice versa (see also Fig.~\ref{Spectral_Jitter}(b,c) which shows two representative photoluminescence spectra of the same nanoparticle). Similar correlated time traces have previously been observed on single CdSe/CdS nanodot/nanorod heterostructures \cite{Muller2004,Muller2005} and CdSe/CdS/ZnS multishell nanoparticles\cite{Gomez2006}. However, the description of this effect was in the first case based on the particular elongated geometry, while in the second case the focus lied on the dielectric environment.

Both the PL shift and its line broadening have been discussed by Empedocles and Bawendi \cite{Empedocles1997} in the framework of the quantum confined Stark effect (QCSE), induced by the presence of fluctuating local electric fields. The QCSE causes an energy shift $\Delta E$ in the PL emission, which depends quadratically on the field strength $\mathcal{E}$, 
\begin{equation}
\Delta E = \alpha \mathcal{E}^2 ,
\label{eqn_DeltaE}
\end{equation}
with $\alpha$ being the polarizability of the confined exciton.\cite{LinearTerm}
A randomly time-varying field will thus result in spectral diffusion. On time scales shorter than the experimental integration time, this diffusion will not be resolved but instead manifest itself in an inhomogeneous contribution $\delta F_{Fluc}$ to the linewidth broadening. This explains the observed correlation between the linewidth and the PL energy. Due to the quadratic nature of the Stark shift, PL energies, which are strongly shifted by the electric field $\mathcal{E}$, will be more susceptible to small field variations $\delta\mathcal{E}$ than PL energies, which are near the energy maximum, the apex of the QCSE parabola. The contribution $\delta F_{Fluc}$ directly follows from Eqn.~(\ref{eqn_DeltaE})
\begin{equation}
\delta F_{Fluc} = \delta(\Delta E) = %\frac{\ud (\Delta E)}{\ud \mathcal{E}} \delta\mathcal{E} = 
2 \alpha\mathcal{E}\, \delta\mathcal{E} = 2 \sqrt{\alpha \Delta E} \, \delta\mathcal{E}.
\label{eqn_deltaFFluc}
\end{equation}
This shows how the PL shift $\Delta E$ and the linewidth broadening $\delta F_{Fluc}$ are correlated. 
Empedocles and Bawendi had already surmised\cite{Empedocles1997} that diffusing charge carriers, located near or close to the QD surface, are responsible for the local electric field.
This raises two interesting questions: (1) Are the charge carriers located in the ZnS shell, in the ligands, in the embedding matrix or in between, at the respective interfaces? (2) How will the $1/r$ Coulomb potential affect the fluctuating field, when the diffusion of the charge carrier will lead to a fluctuating distance $r$? The experiments in Ref. \citenum{Empedocles1997} could be well accounted for, by assuming a constant field variation $\delta\mathcal{E}$ and using Eqn.~(\ref{eqn_deltaFFluc}), which leads to a square root dependence $\delta F_{Fluc} \propto \Delta E^{1/2}$. This is a reasonable assumption for an externally applied electric field, but for the present study, which was conducted at room temperature and in which the Stark shift is induced by fluctuating charges, this assumption is no longer valid. Indeed, plotting the linewidth as a function of the redshift, we observe a superlinear dependence, as shown in Fig.~\ref{Energy_and_FWHM_110908_1_Big}(b). We have therefore developed a model for the line broadening, which takes into account the $r$-dependence of the Coulomb field (see insets in Fig.~\ref{Spectral_Jitter}(b,c)). We start from a randomly diffusing external charge $e$. In a given time interval, determined by the integration time of our detector, it will on average cover a distance $\delta r$. The field fluctuation $\delta \mathcal{E}$ then follows from  $\delta\mathcal{E} = \delta(\frac{e}{4\pi\varepsilon_r\varepsilon_0 r^2}) = 2\sqrt{\frac{4\pi\varepsilon_r\varepsilon_0}{e}} \mathcal{E}^{3/2} \delta r$, which, together with Eqn.~(\ref{eqn_deltaFFluc}), results in 
\begin{equation}
\delta F_{Fluc} = k \left(\Delta E\right)^{5/4} \delta r
\end{equation}
with $k = 4 \sqrt{\frac{4\pi\varepsilon_r\varepsilon_0}{e}} \alpha^{-1/4}$, the CdSe nanoparticle polarizability\cite{Empedocles1997,polarizability}
$\alpha=2.648\cdot 10^{-35}$\,m$^{3}$, and the permittivity\cite{LB_1987} of the CdSe core $\varepsilon_r \approx 10$.

Here, $\delta F_{Fluc}$ is only the contribution to the linewidth broadening, which is induced by fluctuating charges within the nanoparticle's vicinity. It is superimposed upon the linewidth broadening $\delta F_T$ caused by other mechanisms, both homogeneous and inhomogeneous, to give the total linewidth
\begin{align}
FWHM = \delta F &= \sqrt{(\delta F_T)^2 + (\delta F_{Fluc})^2} \\
				 &= \sqrt{(\delta F_T)^2 + k^2(\Delta E)^{5/2} (\delta r)^2}\quad.
				 \label{eqn_DeltaF}
\end{align}

\begin{figure*}[htb]
\includegraphics[width=0.9\textwidth]{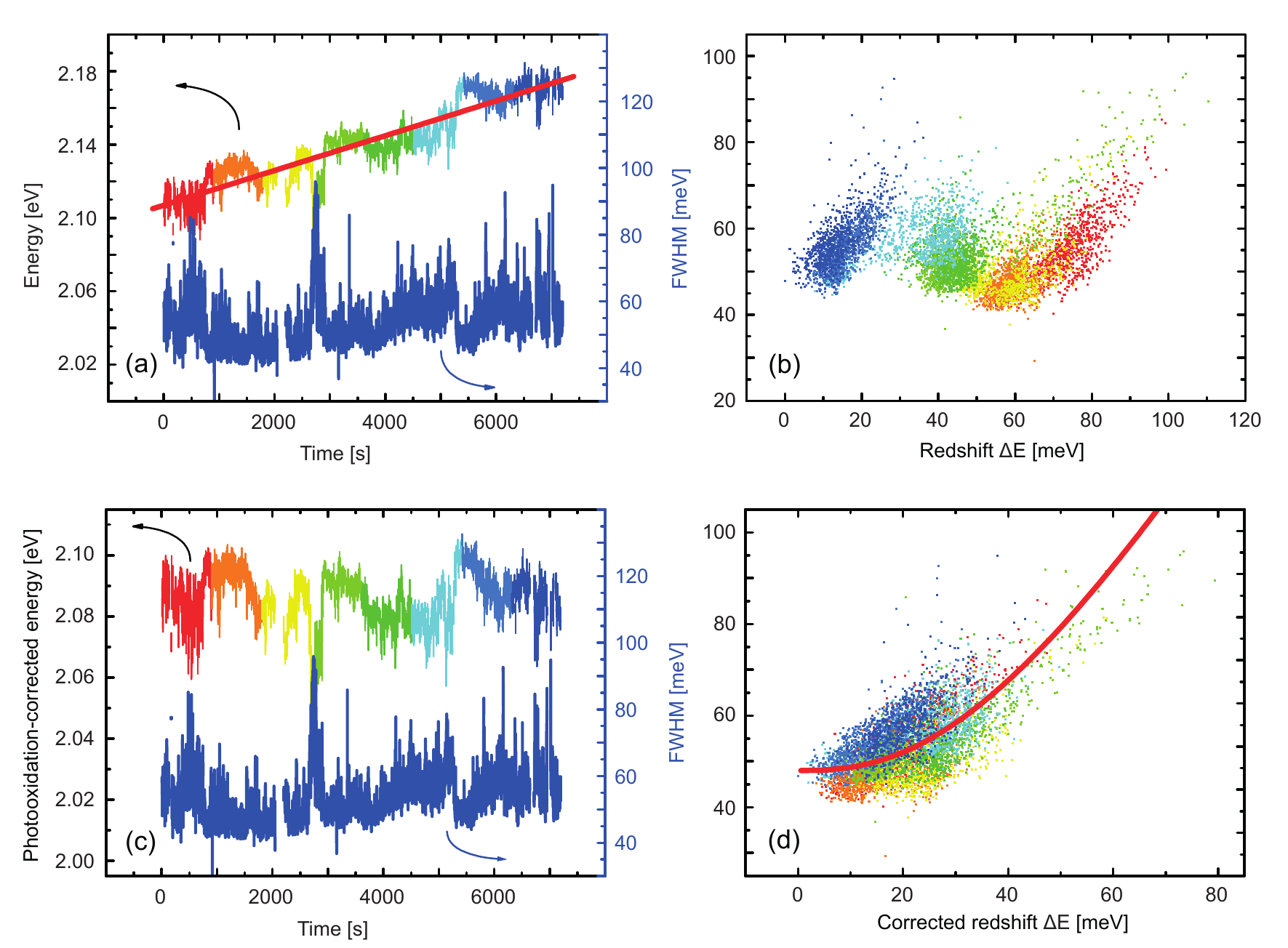}%
   \caption{Spectral diffusion and linewidth broadening of a single CdSe/ZnS nanoparticle influenced by photooxidation, which shifts the emission line towards higher energies ("blueing"). (a) Time-evolution of emission energy (coloured line) and FWHM with a linear fit to approximate the blueing effect. (b) The FWHM in dependence of the redshift for the measurement in (a). (c) Data as in (a), but with the linear energy shift subtracted. (d) Evaluation of the data in (c) as in Fig.~\ref{Energy_and_FWHM_110908_1_Big}(b). The solid line is the same fit with only a slightly adjusted linewidth offset of $\Delta F_T = 49$\,meV.
   \label{Energy_FWHM_Antikorrelation_110921_11_Big2}
   }
%\end{center}
\end{figure*}

The bare linewidth $\delta F_T$ can be easily obtained by noting that at maximum energy applies $\partial E / \partial  \mathcal{E} =0$, so that the fluctuating field will have a vanishing effect on the PL position (see also Eqn.~(\ref{eqn_deltaFFluc})). We find $\delta F_T= \delta F(\Delta E = 0)= 54$\,meV. The average position fluctuation $\delta r$ can also be resolved from a fit to the data. As shown by the solid line in Fig.~\ref{Energy_and_FWHM_110908_1_Big}(b), we find good agreement with the experimental data taking a mean charge carrier displacement of $\delta r = 0.8$\,nm for the given integration time of the detector $\Delta t = 2$\,s.

With Stark's equation $\Delta E = \alpha \mathcal{E}^2$ and the Coulomb field of a single charge, it is also possible to calculate the mean distance of the oscillating charge carrier for a given energy shift $\Delta E$ by
\begin{equation}
\displaystyle r = \sqrt[4]{\frac{\alpha e^2}{(4\pi \varepsilon_r \varepsilon_0)^2 \Delta E}} \quad .
\label{eqn_r}
\end{equation}
Figure~\ref{Energy_and_FWHM_110908_1_Big}(c) shows the statistical distribution of $r$, calculated from all energy shifts in Fig.~\ref{Energy_and_FWHM_110908_1_Big}(b). We find a strongly peaked probability with an average of 3.3\,nm. 

This can directly be compared with the particle dimensions. According to the manufacturer's data, the CdSe core radius is 2\,nm, and the shell and ligand layers have a thickness of 1\,nm and 2\,nm, respectively. These dimensions are confirmed by the PL emission energy and the overall radius of 5\,nm, obtained from transmission electron microscopy images of close-packed particle layers (see inset in Fig.~\ref{Energy_and_FWHM_110908_1_Big}(c)). We therefore conclude that the diffusing charge is located within the ligand layer, ranging from 3 to 5\,nm. This conclusion is in agreement with a number of observations and findings by us and other authors. First, the brush-like structure of the ligands may be much more open to charge diffusion than both the crystalline CdSe/ZnS and the PMMA matrix. The confinement of the charge in the ligands is also in agreement with the findings by Gomez et al.\cite{Gomez2006}, who have found that the PL jitter is independent of the dielectric constant of the embedding polymer. It furthermore explains the findings of several authors \cite{Empedocles1997,Muller2004,Muller2005} that the  charge diffusion is taking place at or near to the particle surface. Finally, the diffusing charge in the ligand layer might in part be responsible for the improvement of the optical properties of CdSe nanoparticles, when the shell thickness is increased \cite{Chen2008}. It should be pointed out that both an inner cut-off (the CdSe/ZnS particle) and an outer cut-off (the PMMA matrix) are necessary to explain the statistics in Fig.~\ref{Energy_and_FWHM_110908_1_Big}(c). Free charge carrier movement would result in an outdiffusion towards infinity, as confirmed by density-of-state considerations as well as numerical simulations (not shown here). It should also be mentioned that diffusion only on the particle surface (at $r = 3$\,nm) cannot explain our data, since we find values up to $r = 5$\,nm, which is 3 times as far from the nanoparticle core as the thickness of the ZnS shell would allow. 82\% of the data are in the range $r \geq 3$\,nm. 

The conclusion that the charge diffusion takes place outside the crystalline particle is further confirmed by PL measurements on particles, which are affected by photooxidation. Figure~\ref{Energy_FWHM_Antikorrelation_110921_11_Big2}(a) shows data taken under ambient conditions. Here, we observe an additional monotonic shift over time towards higher energies, which is not accompanied by a corresponding change in linewidth. During the total observation time of two hours, the PL emission energy increases by about 70\,meV, and shifts of up to 150\,meV have been observed on other samples. Measurements on different particles consistently show that this continuous blue-shift, sometimes referred to as \emph{blueing}\cite{Sark2002,Ito2008,Lee2009}, is irreversible. It is accompanied by a slowly decreasing photoluminescence intensity (not shown here), until the particles become bleached and the PL vanishes. Note that the time scales for both blueing and bleaching are much slower here than those found in literature, in which the nanoparticles usually turn dark within a few minutes or less \cite{Sark2002,Ito2008}. This can be explained by the fact that an illumination power of about 20\,W/cm$^2$ is used here, which is orders of magnitude smaller than in the other experiments.

In addition to the blueing, the above described charge-induced fluctuations also seem to be present in the data. However, a possible correlation between the energy and the linewidth is masked by the continuous blue shift, as seen in Fig.~\ref{Energy_FWHM_Antikorrelation_110921_11_Big2}(b). To indicate the progression in time, the data points have been colour-coded as in Fig.~\ref{Energy_FWHM_Antikorrelation_110921_11_Big2}(a), starting from red around $t = 0$\,s to blue around $t = 7200$\,s. For each colour (or time-slot), a correlation can be discerned in Fig.~\ref{Energy_FWHM_Antikorrelation_110921_11_Big2}(b). To substantiate this, we subtract a linear shift over time (solid line in Fig.~\ref{Energy_FWHM_Antikorrelation_110921_11_Big2}(a) representing the blueing) from the original data, see Fig.~\ref{Energy_FWHM_Antikorrelation_110921_11_Big2}(c). In this corrected data set, the correlation becomes more evident and the data from different time slots become congruent as shown in Fig.~\ref{Energy_FWHM_Antikorrelation_110921_11_Big2}(d). The solid curve shows the same curve of Fig.~\ref{Energy_and_FWHM_110908_1_Big}(b) in which no blueing took place, only with an adjusted linewidth offset $\Delta F_T = 49$\,meV, illustrating the good agreement with our model for this simple linear approach.

The continuous blueshift under illumination can be explained by photooxidation of the inner CdSe core, which reduces its size by up to 1\,nm for nanoparticles of our size distribution \cite{Sark2002}. The resulting increased quantum confinement shifts the emission to shorter wavelengths. As in Ref. \citenum{Ozasa2007}, we find that the blueing can be completely suppressed by keeping the particle under vacuum. A sufficiently thick layer of PMMA works as well.
Of all our nanoparticles measured in vacuum, not a single one showed blueing, but about half of those coated in PMMA and measured under ambient conditions did. Due to a random distribution of the nanoparticles inside the embedding PMMA layer, their protection against the surrounding differs from  particle to particle. Our findings suggest that the particles which exhibit stronger blueing are insufficiently covered by PMMA and can thus interact with the air's oxygen. It is important to notice that photooxidation in general is \emph{not} linear, as experimental data with the other nanoparticles clearly show. But higher order fits, appropriate to the particular particle, lead to the same results as those presented here.

The fact that the blueing and the random fluctuations can be well separated (Fig.~\ref{Energy_FWHM_Antikorrelation_110921_11_Big2}) and that the PL energy/linewidth correlation is not affected by the blueing shows that the dynamics of the fluctuating charge and its influence on the nanoparticle is independent of the oxidation of the particle itself. This further supports our finding that the migrating charge, which causes the PL jitter and the linewidth fluctuations is located inside the ligands. 

%Conclusion
\section{Conclusion}
In summary, we have studied the photoluminescence of single CdSe/ZnS nanoparticles at room temperature and evaluated a well-known correlation between PL energy and linewidth. We find that our data can be well described using a model of a single migrating charge in the vicinity of the particle. Our data allows us to deduce the typical distance of the external charge as a function of the PL energy shift. The deduced values show that the fluctuating charge is located within the ligands layer, surrounding the particles. This conclusion is supported by an evaluation of PL data from particles which also show blueing, caused by photooxidation. The data furthermore allows us to estimate the mean spatial displacement of $\delta r = 0.8$\,nm in the time interval of $\Delta t = 2$\,s.

Our findings show that both the shell thickness and the choice of ligands for CdSe nanoparticles may be crucial steps to reduce spectral jitter and inhomogeneous line broadening, two effects, which are commonly found in nanoparticle devices, particularly when working at technically relevant temperatures. Therefore, more attention should be given to the ligand layer when trying to improve the optical properties of nanoparticle-based devices.

%\begin{acknowledgement}
%The authors thank the workgroup for fruitful discussion.
%\end{acknowledgement}
 
 \bibliography{The_role_of_the_ligand_layer}

\begin{thebibliography}{33}
\bibitem [{Lin()}]{LinearTerm}%
  \BibitemOpen
  \href@noop {} {}\bibinfo {note} {For simplicity, we have eliminated a linear
  term from a possible permanent dipole in Eqn.~(\ref{eqn_DeltaE}), by
  rescaling the energy and the electric field to the apex of the
  parabola.}\BibitemShut {Stop}%
\bibitem [{pol()}]{polarizability}%
  \BibitemOpen
  \href@noop {} {}\bibinfo {note} {It is important to note that there is an
  inconsistency in the definition of $\alpha$ in the literature due to the use
  of external or internal fields, mismatching the polarizability by a factor of
  $\varepsilon_r^2$. Electron and hole of the exciton are pulled apart by the
  effective internal electric field, thus both our calculations and the taken
  polarizability are based consistently on internal fields, which take the
  permittivity into account.}\BibitemShut {Stop}%
\end{thebibliography}
 
\end{document}